\documentstyle[psfig]{l-school}

\begin{document}
\markboth{L. Tresse}{LUMINOSITY FUNCTIONS AND EVOLUTION}
\setcounter{part}{10}
\title{LUMINOSITY FUNCTIONS AND FIELD GALAXY POPULATION EVOLUTION}
\author{L. Tresse}
\institute{
Istituto di Radioastronomia del CNR, Via Gobetti, 101, I-40129 Bologna \\
Laboratoire d'Astronomie Spatiale, BP8, F-13376 Marseille Cedex 12}
\maketitle

\def\mvup{\phantom{a}\vspace*{-5mm}} 
\parskip 0pt

\section{INTRODUCTION}
Galaxy redshift surveys are outstanding tools for observational
cosmology.  Mapping the universe as outlined by galaxies leads to
fundamental measurements which refine our knowledge of its structure
and evolution.  Redshift acquisition has undergone tremendous progress
thanks to advances in technology, and redshift surveys appear nowadays
as routine.  Even though they may look simple at face-value, the
strategy of a survey and the galaxy-selection criteria have crucial
impacts on the interpretation of results.  Since galaxies are directly
observable point-like tracers of dark matter halos, they represent
only the tip of the iceberg of what drives the evolution of the
universe.  Hence interpretation of these surveys via model-dependent
approaches such as semi-analytical models, N-body simulations provide
also fundamental insights into galaxy evolution and formation.
Redshift surveys can be analyzed using many different statistical
techniques to give measurements of clustering, large-scale structures,
velocity fields, luminosity functions, weak-lensing.  

In this lecture, I concentrate on one of these measurements, i.e. the
galaxy luminosity function (LF) and its evolution. In particular I
consider the LF derived from optically magnitude-selected field galaxy
redshift surveys.  The LF is a fundamental measurement of the
statistical properties of the population of galaxies; it is the
comoving number density of galaxies as a function of their intrinsic
luminosity.  The luminosity of a galaxy evolves according to the
evolution of its content (see lectures of Charlot and Matteucci), and
according to its interaction/mass accretion/merging history.
Measuring the LF at different cosmic epochs enables us to quantify its
changes and to assess the evolution in the galaxy population.  Early
work on galaxy number counts as a test for evolution has largely been
superseded by large redshift surveys which allow the direct
determination of the LF from which the observed N($m$) and N($z$) can
be reproduced and better understood.  Ultra deep number counts are
still used at depths where automatic redshift acquisition is not yet
possible (see Ferguson's lecture for an up-to-date review of number
counts).  I discuss only field galaxies which are selected without
regard to their environment or any special properties, hence their
study is representative of the global galaxy population.  LFs have
also been measured from samples which select only either AGN galaxies,
or radio galaxies, or H$\alpha$ emitters, or galaxies in clusters, or
clusters, etc.  These latter measurements determine the evolution of
single populations; but the completely different selection means that
their connection to the field galaxy population is not
straightforward. With large, distant and multi-wavelength surveys of
field galaxies, it will be possible to measure the LF for the global
population, and for each of these special subsamples, and thus to
relate them more easily at different epochs.
 
The aim of this lecture is to describe the approach to designing
state-of-the-art of galaxy surveys and their impact on measurements on
LFs.  I start by discussing survey strategies, which I consider to be
the most important step (Section~2). I continue by describing the data
required (Section~3), and then review the different estimators to
measure LFs (Section~4). I discuss the LF evolution (Section~5) and I
follow by summarizing the status of LF measurements (Section~6).  I
would like to emphasize that a survey for which one can define and
quantify the biases and selection criteria, is more useful in the long
term than a survey for which they are not well determined, no matter
how pioneering the work is.  Indeed comparing data from different
surveys is a nightmare, and comparing them to models is even less
straightforward. All surveys have different selections, biases and
methodologies, and so the interpretation of apparent discrepancies
should be done with care before invoking any exotic explanations.  The
references quoted for the surveys are usually those estimating the LF;
references for the series of papers issued from a survey should be
found within.

\mvup
\vspace{-0.4cm}
\section{SURVEY STRATEGIES}
A survey strategy includes choices and constraints at several levels,
some of which are described in this section.  If not well-defined and
well-controlled, poor choices can produce unknown biases toward
certain types of galaxy, which will hinder and confuse analysis of the
survey.

\subsection{From the local to deep universe}  
One crucial aspect of the preparation of a redshift survey is to find
a balance between the sky coverage (i.e. the area of sky in which
galaxies are selected for spectroscopic observations), the sampling
rate (i.e. the number of spectroscopic targets out of the number
of photometric objects within a magnitude range) and reasonable
observing times (which depend on the detectors and telescopes used).
If these are not handled, it leads to major difficulties to interpret
and compare LFs, and even worse for correlation functions or
close-pair analysis.  Observational strategies are quite different for
the local universe ($z\!<\!0.1$) and the distant universe ($z\!\gg\!
0.1$).  

What matters for the local universe ($z\!<\!0.1$) is the large sky
coverage, indeed any single small area is strongly affected by density
inhomogeneities.  Until recently the way to survey large areas was to
use photographic plates of $\sim$25 sq. deg field of view.  The SDSS
(\cite{lovsloan}) will be the first large local survey with galaxies
selected from CCDs.  For local surveys the magnitude limit is about
$B\!\sim$16, and exposure time required for a spectrum is a few
minutes.  A full sampling strategy is necessary when aiming to refine
the local structures (CfA2 \cite{mar94}, SSRS2 \cite{dac}), see da
Costa's lecture), but is time consuming (over several years, changes
in detectors, strategies lead to disparities within one sample).
Meanwhile sparse-sampling strategies have also been adopted (SAPM
\cite{lov92}), or a collection of pencil-beams (KOSS \cite{kir1}).
While these previous surveys observed spectra one by one, wide-field
multi-object spectroscopy has been more recently used by Autofib/LDSS
\cite{ell}, LCRS \cite{lin}, CS \cite{gel}, and ESP \cite{zuc}
surveys. The on-going 2dF (\cite{mad}) and SDSS use even larger
multiplex gain, respectively 2x200, and 2x320 fibers in one exposure
($\sim$1 hour). Sampling galaxies $B\!\simeq$19.5, these will have a
mean depth of $z\!\simeq$0.1.
 
Pushing deeper ($0.2\!<\!z\!<\!1.3$) and with multi-object
spectroscopy, redshift surveys (see references in Table~2) are done in
pencil beams of few arcmins, with faint magnitude limit ($B\!<\!24$)
and long exposure spectra (few hours).  The acquisition of several
pencil beams is necessary either to cover contiguous patches on the
sky or to sample different fields of view over the sky, and several
exposures are usually required on the same field.  One of the original
aim of these surveys was the LFs (and also clustering), indeed
studying structures is reliable only with larger areas.  Surveys at
$1.3\!<\!z\!<\!2.7$ require detectors sensitive in the infrared since
no strong spectral features are observed in the optical wavelength
range. This cosmic epoch will very soon be surveyed by 8m class
telescope (see lecture of Le F\`evre). For higher redshifts
($z\!<\!4-5$), the existing redshift survey (\cite{stei}) has selected
only a certain type of galaxies (the Lyman break population).
Systematic redshift surveys in this range will be also conducted with
8m telescopes. The universe at $z\!>\!5$ is still an unknown regime,
but some models suggest that there is a population of galaxies already
formed.  If such a population exists, the distances between galaxies
(or any subgalactic mass undergoing star formation) will be smaller,
and the apparent size of galaxies larger, leading to very crowded
fields. Thus methods such as Integral Field Unit may be chosen rather
than traditional MOS, and are under studies in particular for the Next
Generation Space Telescope as well as Micro-Mirror or Micro-Shutter
Arrays MOS (see ``NGST - Science Drivers \& Technological Challenges'',
34th Liege International Astrophysics Colloquium, June 1998, ESA
Publications).

\subsection{MOS modes}
Each survey using multi-object spectroscopy (MOS) has biases
introduced by the constraints inherent to the particular MOS mode
used.  For instance when using fiber multiplex, a set of fibers is
placed on objects while another set is placed on sky.  Since the sky
spectrum is measured through different fibres than the galaxies, the
throughput and spectral response will be different and this can lead
to poor sky subtraction, especially for faint galaxies whose flux is a
small fraction of sky.  Also fibres typically cannot be placed closer
than $20''$ from one another, which leads to a bias against close
pairs of galaxies unless fields are observed several times.  Using
slit multiplex, a slit is placed on an object including sky on both
side, and can include close objects.  However, while the output
spectrum from a set of fibers is arbitrarily reorganized on the CCD,
it is not the case with slits for which the spectrum is dispersed on
both sides of the aperture location on the CCD. No other objects can
be observed in the CCD area covered by the spectrum within one
observation.  This produces non-uniform selection patterns in the
dispersion direction, i.e. some areas are more sampled than
others. Depending on the number of observations (or the sampling rate
aimed), these effects can be reduced but at the cost of the volume
surveyed within a certain allocated time.  Basically the number of
possible slits/fibers to be positioned on galaxies is function of
galaxy density (related to the magnitude range sampled) and the number
of observations on the same sky area.  Thus to get a 100\% sampling is
somewhat time consuming. A compromise has to be reached between the
full use of the multiplex gain and the uniformity in ($x$, $y$) of the
galaxies selected.  There are even more subtle biases with both MOS
modes. The quantification of such effects is not straightforward;
simulations are helpful to measure them, and take them into account in
statistical analysis.  

\subsection{A priori selections}
One common pre-selection for objects to be observed in spectroscopy
mode is the galaxy/star separation.  Depending on the survey magnitude
range, the number of stars can be so high that it is necessary to do
this pre-selection to avoid spending most of the time observing
stars. For instance in the SAPM magnitude range ($15\!<\!b_J\!<\!17$,
see \cite{lov96}), stars are $\sim$95\% of the objects!  However the
galaxy/star separation techniques are not 100\% reliable, for example
such pre-selection is likely to exclude compact galaxies (in the APM
it has been estimated at 5\%, \cite{mad90}).  For deep surveys this is
less crucial.  For instance in the CFRS magnitude range
($17\!<\!I\!<\!22$) no-preselection was done, and depending on the
field galactic latitudes, from 10 to 30\% of objects were stars of
mostly M and K types. Another pre-selection is to choose the objects
by eye. In these cases, nice objects, close objects, peculiar objects
will tend to be selected introducing a non-controlled bias.

\subsection{Photometric choices}
Objects in redshift surveys were usually selected from one pass-band
of wavelengths, and this leads to a preference for a certain type of
galaxy.  Since the detectors were the most sensitive in the blue, the
first redshift surveys selected galaxies from their flux received
around 4400 \AA.  Thus the observed rest-frame luminosity spans the
ultraviolet wavelengths ($4400/(1\!+\!z)$), and this comes down to
select galaxies mostly on their current star-formation activity.  The
UV emission is known to be strongly dust extinguished (for instance
the flux is reduced by a factor $2-3$ for galaxies at $z\!\sim$0.2,
\cite{tremad}), and is usually much fainter than the optical for
spiral and elliptical galaxies. Hence galaxies become difficult to
detect especially at $z\!>\!0.5$, and those observed are likely to be
only the strong star-forming galaxy population. Recent deep surveys
selected galaxies from their red or infrared observed flux. This
enables us to observe galaxies at $z\!>\!0.5$ more easily since the
spanned rest-frame light lies in the optical. This selection is closer
to the galaxy mass regardless the current star-forming activity, and
so the observed population is more representative of the total galaxy
population. 
 
Single pass-band surveys select galaxy populations which are never
fully homogeneous from one cosmic epoch to another. To quantify the
fraction and type of galaxies over or under-observed is not
straightforward, and requires multi-color selected samples. This leads
to LF measurements at various epochs which are derived from different
observed galaxy populations, and it has to be handled with care when
studying evolution of LF.  Color-selected samples searching for the
Lyman break (or UV dropout), are recently used for galaxies at
$z\!>\!3$, but select only the galaxy population for which this break
is detectable (\cite{stei}).  With large on-going multi-wavelength
surveys, several samples of multi-color selection should be more
homogeneous.
 
Another important selection criterion is the way the magnitudes are
measured, and thus how the objects are selected.  Photographic
magnitudes are more difficult to calibrate than CCD magnitudes, due to
the non-linearity and saturation occurring in photographic plates;
these effects are non trivial to correct, even when calibrating with a
sub-set of CCD magnitudes.  To be complete down to a certain magnitude
in a single band, the best is to use total magnitudes.  However since
the outer part of galaxies is usually below the threshold detection
(i.e. a small fraction of the surface brightness of the night sky), 
in practice total magnitudes are not measured directly and are
retrieved from aperture or isophotal magnitudes.  Recovering for the
lost flux can be subject to systematic errors and ultimately requires
knowledge of the surface brightness (SB) profile, and thus most
surveys quote a photometric completeness based on their choice of flux
measurement. 

Actually catalogues are limited by an apparent magnitude, a SB and a
size. The usual truncation of a catalogue to a magnitude limit results
in biases in SB and size which add complexity when comparing surveys,
or when recovering the galaxy densities (see for instance
\cite{pet98}, \cite{impey} and references within).  Isophotal
magnitudes include a varying fraction of the total flux due the
($1+z$)$^{-4}$ SB dimming effect, and different galaxy SB profiles; to
be close to the total magnitude the adopted isophote limit must
correspond to about 6 magnitudes fainter than the total magnitude (for
instance if m$_{tot}$~=~22, SB~=~28 mag arcsec$^{-2}$, \cite{lil1}).
Petrosian magnitudes integrate the flux within a radius determined by
the ratio of the mean SB within radius $r$ to the local surface
brightness at $r$ (\cite{pet}).  Aperture magnitudes can be used with
an empirical fit to the galaxy light-growth curve to integrate the
flux within an area close to the total flux emitting area; the
required galaxy image center is usually recovered from an isophotal
measurement.  For instance Kron aperture magnitudes (\cite{kron80})
integrate the flux within a radius which is a multiple of the first
moment radius ($r_1$) of the intensity-weighted radial profile
($r_{Kron}\simeq 2 r_1$).  The integrated flux with these methods
reliably reaches 90-95\% of the total flux.  Some complications occur
when galaxy images overlap, or when strong emitting regions in the
outer disk can be taken mistaken for another object.

To be complete down to a certain color, the light must be integrated
from the same emitting area of a galaxy. This can be done within a
fixed aperture and considering the same center for different pass-band
images, or using an isophotal area defined from the sum of these
images.  Magnitudes that are close to total magnitudes are the most
attractive to use in defining a sample, because it is then easy to
select well-defined sub-samples as, for instance, a fixed aperture
magnitude sample for work on a color-selected sample.  Starting from a
fixed aperture-selected sample it is not possible to generate a sample
complete to a given total magnitude.  Total magnitudes are also
attractive to study low-surface brightness and intrinsically faint
galaxies.

\subsection{Spectroscopic choices} 
The choice of the spectral resolution ($\Delta\lambda/\lambda$) has an
impact on the success rate for measuring a $z$.  For instance, if the
observed $z$ range (related to the magnitude range) is large, a
low-resolution allows to observe a large wavelength range, and
increases the chances to observe an emission line as the common
[O~{\sc ii}], and/or the Balmer break.  On the contrary, a
high-resolution is better to work out the spectral properties of
galaxies, including H$\delta$, H$\beta$, $d_{4000}$, H$\alpha$,
[O~{\sc ii}] measurements for star-formation rate, metallicity and
spectral classification.  With new infrared detectors to observe
galaxies at $1\!<\!z\!<\!3$, a high resolution is better to correct
for the strong OH sky emission and assure the $z$ measurement.
Another point is that it is preferable to span wavelengths that are
within the filter-band of selection for galaxies so that the
spectroscopy completeness correlates with the photometric
observations.  To measure a $z$, flux calibration is not
necessary. However again if the spectral properties are studied at
several $z$, it is better to obtain calibrated spectra.
Spectro-photometric calibration also enables us to correct for
aperture light losses.  Several exposures are required to filter out
the cosmics.  A signal-to-noise of 10 or more is required if one
wants to avoid a bias against spectra with weak features.  The choice
of the aperture has also its importance in the success rate for
finding a redshift; slits of $\sim\!1.5''$ are usually used for deep
surveys. The success rates for redshift surveys can reach levels of
$\sim$90\%, and are usually function of the magnitude; at the faint
limit of a survey incompleteness can be very large!

\section{BASIC DATA REQUIRED TO MEASURE LF} 
The basic data needed for each targeted object of a survey are the
redshift $z$, the relative magnitude $m$, and the galaxy type.  It is
important that each step is well-defined so that the redshift
completeness function and the uncertainties in the absolute magnitude
determination can be established for the survey.

\subsection{Redshifts}
The method to measure $z$ from a sky-subtracted optical spectrum is to
identify a set of spectral features which have the same
($\lambda_{obs}/\lambda_{rest} = 1\!+\!z$).  Note that
$\lambda_{rest}\!>\!2000$\AA\ is measured in the air, while
$\lambda_{rest}\!<\!2000$\AA\ in the vacuum; in this last case $(1+z)$
has to be multiplied by the refractive index of air ($n=1.0029$).  The
reliability of $z$ is correlated to the number of features and their
signal-to-noise; for instance a single strong emission line is not
sufficient to determine securely a $z$, also for a set of several weak
features (for these cases, colors may be helpful to secure $z$, see
below).  The simplest way to measure $z$ is to fit individual lines,
and/or to cross-correlate with template spectra.  Now that thousands
of spectra are taken by night, fully automatic measurements of $z$ are
necessary; hence supplementary procedures have been developed using
for instance Principal Component Analysis (PCA) techniques (see
e.g. \cite{con}).  The difficult task of such automation is to account
for all possible situations as for instance the diversity of spectra
(from extremely blue to red spectra, from QSO to stellar spectra, from
featureless spectra to very reddened spectra), the problems in the
spectra (bad sky subtraction, etc.).  Such things could be easily
judged by eye, but with the large on-going and future surveys it is
impractical to look at them one by one.

In the optical window [4000--9000]\AA, the strongest features are
[O~{\sc ii}] $\lambda$3727, the Balmer break at 4000\AA, H$\beta$,
[O~{\sc iii}] $\lambda\lambda$4959,5007, H$\alpha$, [N~{\sc ii}]
$\lambda\lambda$6548,6583, [S~{\sc ii}] $\lambda\lambda$6718,6731, and
followed by H$\gamma$, H$\delta$, CaH, CaK.  With [O~{\sc ii}] in the
optical, one can determine $z$ until 1.4 about, then no strong
features appear in the optical until $z\!\sim$2.2 where Ly$\alpha$
(1215\AA) can be observed. To cover the missing region, detectors
sensitive in the near infrared window [0.9--1.5]$\mu$m are required to
still observe at least [O~{\sc ii}]. 
 
In principle, measured redshifts must be transformed from the Earth to
the Sun system (heliocentric $z$) then to the galactic center
(galactocentric $z$).  This represents about 300 km s$^{-1}$ or 0.001
in $z$.  In the very nearby universe ($z\!<\!0.03$), peculiar
velocities of galaxies are not negligible compare to the Hubble radial
velocity flow of 300--500 km s$^{-1}$, i.e. when $v_{radial} < 10 000$
km s$^{-1}$.  The accuracy usually aimed is better than 0.001 in $z$.
Another point is that absorption lines do not always give the same $z$
as emission lines (velocities can differ by up to $\sim$500 km
s$^{-1}$).  This is usually explained by the fact that emission lines
originate in different regions than absorption lines, and that this
effect is enhanced when galaxies are seen edge-on. 
 
Another way to determine redshifts is to use the colors of a galaxy,
which corresponds to using an extremely poor resolution spectrum since
a color is the integrated flux over several wavelengths. Thus they are
inferred from the spectrum shape, and mainly from the location of the
discontinuities in the continuum (Balmer, Lyman breaks).  These
redshifts are called photometric redshifts; they rely strongly on the
knowledge of the spectral energy distribution and its evolution for
any type of galaxies, and cannot give precise measurement.  To avoid
catastrophic identifications, UV and IR colors are absolutely
essential, and the error in $z$ is reduced to $\sim$0.03 in $z$
(\cite{con95}).  They were first used when acquisition of spectra was
extremely time consuming, and have recently came back to fashion to
estimate redshifts of very distant galaxies as seen in the Hubble Deep
Field, where spectroscopy requires at least 8m telescopes.  They are
particularly useful in the presently region at $z\!=\!1.4-2.2$, where
there are no strong optical features.  Actually they are a useful
technique to estimate the $z$ of a galaxy, and thus to select the
window (optical or infrared, thus the instrument) in which prominent
features are likely to be seen in spectroscopy.

If no redshift can be measured for a galaxy, then it is part of the
redshift incompleteness of the survey.  It is important to quantify
this as a function of magnitude, and to understand where it may come
from (only low SB galaxies?, only high-$z$ galaxies?, problem with the
slit/fiber, etc.).

\subsection{Absolute magnitudes}
\begin{figure}
\centerline{\psfig{figure=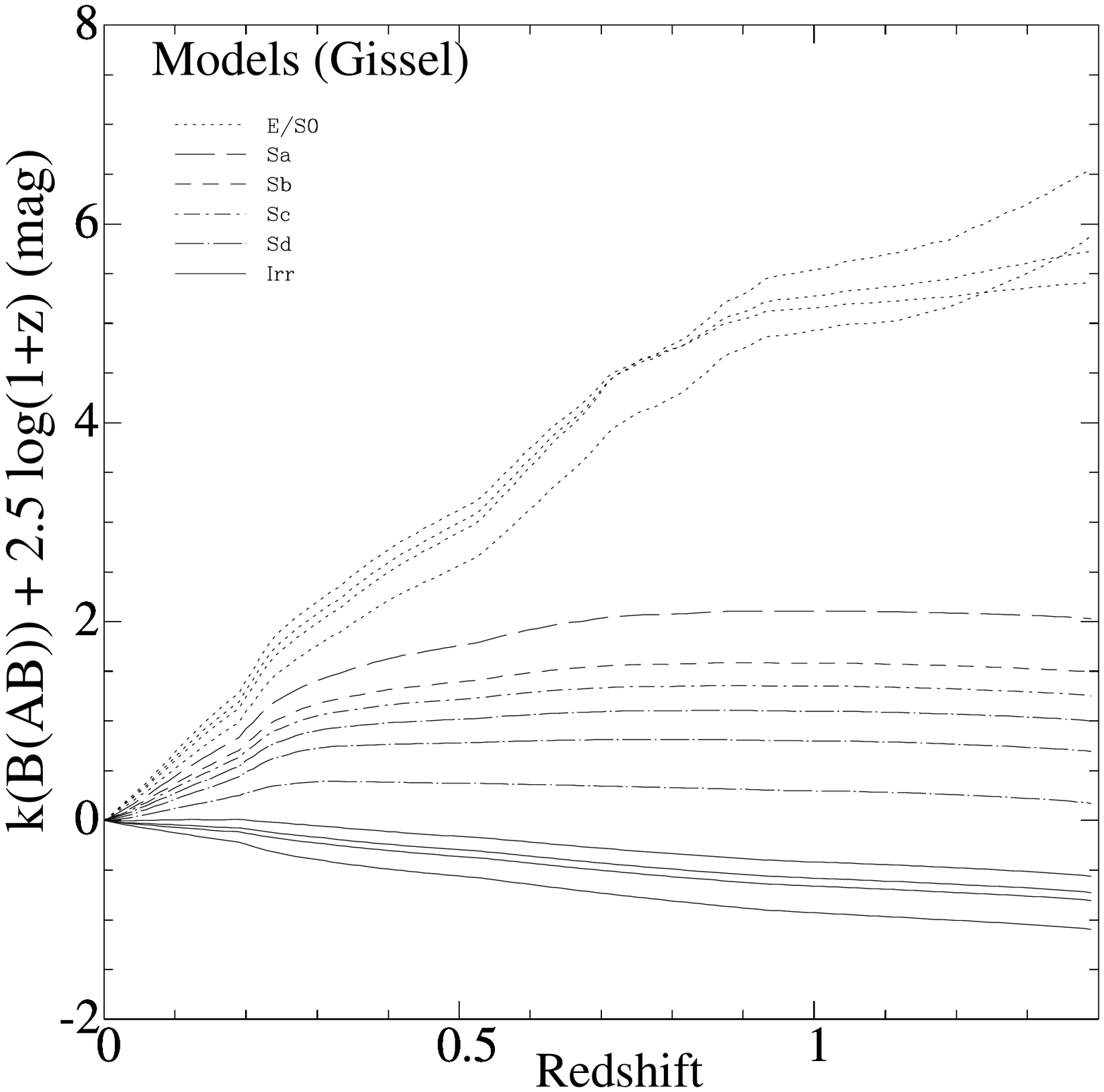,height=6cm}
\psfig{figure=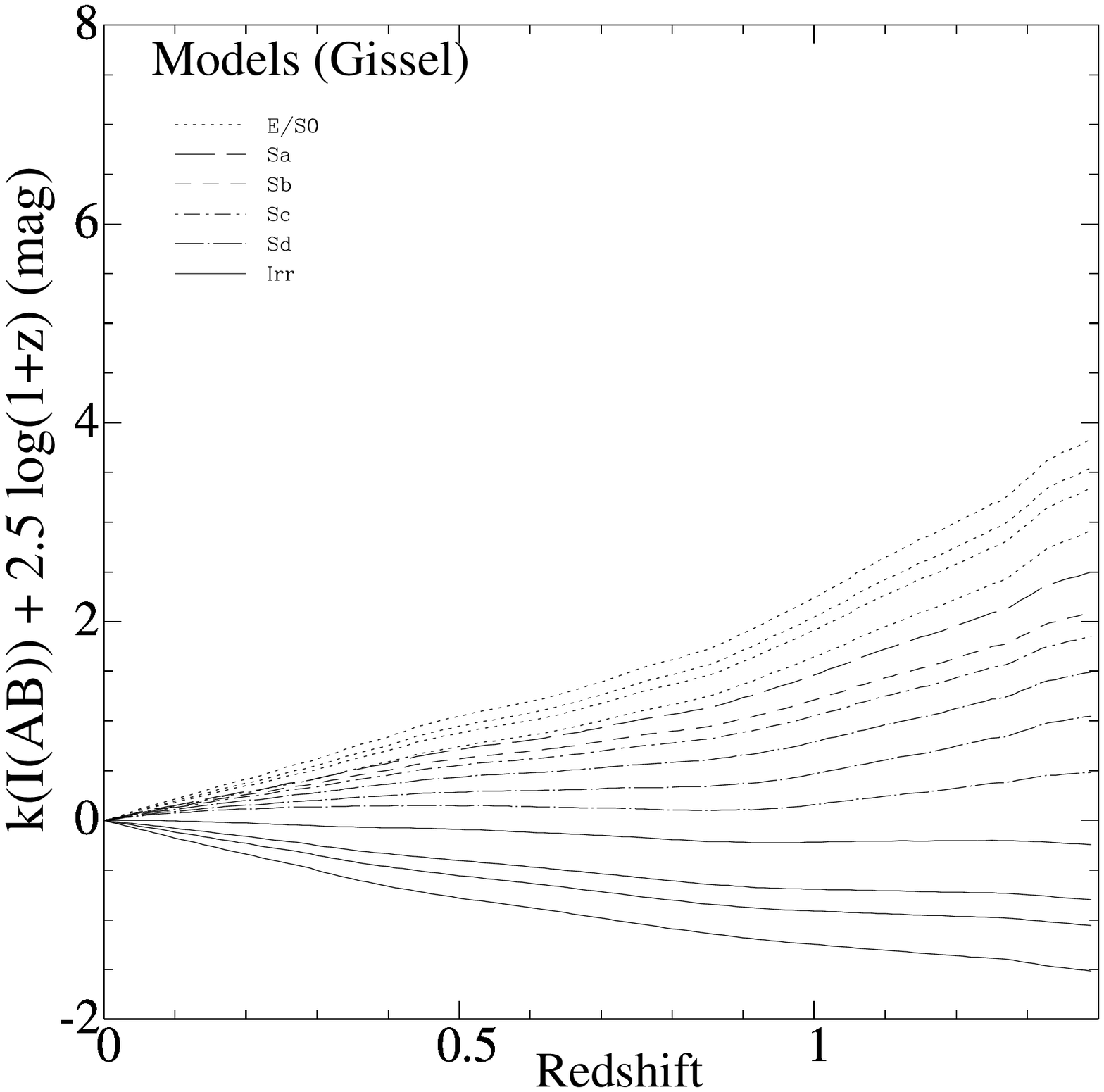,height=6cm}}
\vspace{-0.25cm}
\caption[fig1]{$k$-corrections minus $2.5\log(1\!+\!z)$ in B$_{AB}$ 
(left) and in I$_{AB}$ (right) using Gissel types (see \cite{bru})
from E (top lines) to Irr (bottom lines) spectral types. \label{fig1}}
\end{figure}

Magnitudes are determined in a well-defined pass-band such as $B, V,
R, I$, or $K$. Denoting the band as $j$, the absolute magnitude is
$M_j=m_j-DM-k_j-A_j$ where $DM$ is the modulus distance depending only
on the world model chosen (H$_{0}$ and q$_{0}$), $m$ the apparent
magnitude, $k$ the correction necessary to express all magnitudes in
the same rest-frame filter band, and $A$ the galactic extinction.
This is the classical way to measure absolute magnitudes in redshift
surveys, which minimizes model-dependent inputs.  Then when comparing
with models, these magnitudes can be corrected, as wished, for the
predicted luminosity evolution of a certain type of galaxy at $z$ in
the same rest-frame band, and for reddening produced by the dust
extinction intrinsic at each galaxy.  However accounting only for the
$k$-correction, the measure of $M$ is already model-dependent
(assuming a world model).  Depending on the galaxy type the
model-dependent term of $k$-corrections spans $\sim$4 magnitudes in B,
and $\sim$2 magnitudes in I at $z\!=\!1$ (see Fig.~\ref{fig1}).  Thus
it is crucial to determine accurately the galaxy type.  The best is to
use (multi-)color information; the spectral continuum may also be used
but sometimes cannot accurately constrain a type depending on the
rest-frame wavelength range observed and the spectral resolution.
Morphological information has also been used for local galaxies, but
because of the miss- or non-classification of certain galaxies it is
not as reliable as colors.  One way to minimize the $k$-correction is
to use the relative magnitude which spans the rest-frame band in which
absolute magnitudes are expressed. For instance, for galaxies around
$z\!\sim$0.2, $\sim$0.9 and $\sim$2.7, respectively m$_V$, m$_I$ and
m$_H$ spans the rest-frame B band, so the $k$-correction is small, and
at these exact $z$, the model-dependent term of the $k$-correction is
null.  Also $k$-correction can be directly measured from a
flux-calibrated spectrum if the observed wavelength range includes
both the observed- and rest-frame bands; but this measure depends on
the quality of the calibrated spectra.
The proper determination of $k$-corrections has an impact on the
accuracy of $M$ and on the measure of the observable volume of a
galaxy used in the LF estimators.

\section{ESTIMATORS FOR THE LUMINOSITY FUNCTION}
Several methods have been developed and generalized to estimate the
comoving number density as a function of luminosity, i.e. the {\it
luminosity function} (LF) $\phi(L)$ (or $\phi(M)$) expressed as a
number of galaxies per Jansky (or per magnitude) per cubic megaparsec.
Table~1 summarizes the estimators that have been used; for a
comprehensive review on estimators and history, see \cite{will} and
references within.  These estimators usually assume that the
luminosity $L$ is uncorrelated with spatial location ${\bf r}$, so
that the comoving number density at a distance ${\bf r}$ as a function
of luminosity can be written as $\Phi(L, {\bf r}) = (\rho({\bf
r})/\overline{\rho({\bf r})}) \ \phi(L)$ where $\rho({\bf r})$ is the
galaxy {\it density function} (DF).  The assumed separability of LF
and DF means for instance that the LF is supposed to be the same in
clusters and in the field which have different DF. This assumption led
to estimators which are independent of the spatial galaxy distribution
even though it is inhomogeneous on small and large scales.  I
emphasize now that correlations {\em are} observed between galaxy
properties and the neighborhood density.  The separability of $L$ and
${\bf r}$ remains if the LF is calculated at a certain cosmic epoch,
with estimators extended to be applied as a function of redshift.
Hence estimators assume that $\phi(L)$ is uncorrelated to ($x$, $y$),
and changes only a little within a small range of redshifts.
\begin{table}
\caption{Table summarizing different methods to estimate the LF (see
text) with in column (1) the estimator name, (2) the assumption or not
of a parametric function for the LF, (3) the assumption on $\rho({\bf r})$,
(4) the output for the LF, (5) the references which are not exhaustive
for a question of space (fully detailed ones are given in 
[2] and [57]).}
\vspace{-0.5cm}
\small
\begin{tabbing}
\hspace{2cm}\=\hspace{1cm}\=\hspace{2cm}\=\hspace{2.5cm}\=\hspace{2cm}\=\kill
\hrulefill \\ 
(1) \> (2) \> (3)  \> (4) \> (5) \\
\hrulefill \\ 
V$_{max}$   \> no \> uniform$^1$  \>  $\phi$, L, $\alpha$ \> \cite{schm}, \cite{fel}, \cite{eal} \\
C$^{-}$   \>  no \> spherical \> $\phi$, L, $\alpha$ \> \cite{lyn}, \cite{cho} \\
$\phi/\Phi$ \> no \>  none \>  L, $\alpha$ \> \cite{tur}, \cite{kir}, \cite{dav} \\
STY \> yes \> none \>  L, $\alpha$ \> \cite{san} \\
SWML \> no \> none \>  L, $\alpha$ \> \cite{efs}, \cite{hey}, \cite{spri} \\
\hrulefill  \\
$^1$ In practice, the density is assumed constant in the range of redshifts 
in which \\ the  LF is estimated. 
\end{tabbing}
\end{table}
\normalsize

Estimators for the LF differ mainly in how they evaluate the
probability of observing a galaxy with a luminosity $L$ and type $i$
in a given volume of the universe. The volume is usually defined by
the low and high apparent magnitude limits of a survey, and by the
redshift bin analysed giving a minimal and maximal redshift.  Actually
the SB and size limits should also be taken into account if
well-defined samples in SB and size are used. These estimators rely on
the Bayes' theorem which in this case says that, in the absence of
prior information, the most likely LF given the observations is the
one which would most often reproduce the observed distribution of
galaxy luminosities in a series of equally likely realizations.  Thus
we maximize the joint probability for the observables. 
Usually this is done by maximizing the likelihood function, 
defined as $ log_e {\cal L} = log_e \prod_i p(L_i) = \sum_i p(L_i)$.

Briefly, V$_{max}$ estimators are simply based on the inverse sum of
the maximum observable comoving volumes of each galaxy.  They assume a
uniform density and thus are affected by clustering in the galaxy
distribution.  STY, SWML and $\phi/\Phi$ estimators cancel out in the
calculations the density, so they are clustering insensitive methods.
The C$^{-}$ method assumes a spherical density, i.e. the LF has the
same shape at any ($x$, $y$), so is ideal for pencil-beam rather than
large-angle surveys.  We note that all estimators can account for
the completeness function.

Local surveys ($\bar{z}\!\sim$0.05) are strongly affected by the
density fluctuations of the Virgo and Coma clusters in the North
hemisphere, and by a local void in the South.  For example, Willmer
(\cite{will}) compares different estimates of the LF, using the CfA1
redshift survey, and find large discrepancies caused by the Virgo
cluster which dominates this survey. For slightly deeper local surveys
($\bar{z}\!\sim$0.1), these discrepancies are much reduced.  The
V$_{max}$ density-dependent estimates give higher faint-end shape when
the observed area is cluster dominated, but if the area is large and
deep enough, the effects of clusters and voids are counter-balanced
and the V$_{max}$ estimate is similar to clustering insensitive
estimates as seen in the 2dF (\cite{mad}).  Distant surveys
($z\!\gg\!0.1$) which sample galaxies in pencil-beam fields of view do
not suffer from these local inhomogeneities, but their line of sight
through cosmic epochs also crosses structures and voids. Observing
several pencil-beams over the sky smoothes this out, and discrepancies
between estimates are then small (usually within Poisson error bars).

Even though methods do not assume a parametric function for $\phi(L)$
(save the STY method), the shape of LFs are usually fitted with a
Schechter function, i.e.  $\phi(L) dL =\phi^{*} (L/L^{*})^{\alpha}
\exp(-L/L^{*}) d(L/L^{*}), $ where $L^*$ represents a characteristic
luminosity above which the density of more luminous galaxies decreases
exponentially, $\alpha$ is the slope of the LF at fainter luminosities
(called the faint-end slope), and $\phi^{*}$ is the number density of
galaxies at $L^{*}$ (called the normalization). In practice the number
densities per magnitude are fitted, i.e.  $\phi(M) dM = 0.4 log_e(10)
\phi^{*} (10^{-0.4(M-M^*)})^{(\alpha+1)} \exp(-10^{0.4(M^*-M)})
dM$. Thus we say the slope is steep, flat or negative respectively when 
($\alpha+1$)$<$0, $=$0 or $>$0.  The three Schechter parameters are
highly correlated, and thus redshift surveys which sample a large
range of luminosities are likely to give the best results for the LF
estimation. It is of course a challenge to achieve this at any
redshift.  The mean galaxy density and mean luminosity density in a
comoving volume are respectively $ \int \phi(L) dL = \phi^{*}
\Gamma(\alpha\!+\!1)$ and $ \int L \phi(L) dL = \phi^{*} L^{*}
\Gamma(\alpha\!+\!2). $ The slope is measured $\alpha\!<\!-1$; thus
for a non-diverging density the LF must have a cut-off at faint
$L$. This cut-off has not yet been observed even in the deepest local
survey (ESP). Since the slope is also measured $>\!-2$, the numerous
faint $L$ galaxies contribute little to the total luminosity density.

Recently, it appears that the Schechter function is not a good fit to
the overall LF over a wide luminosity range (\cite{lov97},
\cite{spra}, \cite{zuc}). A modified Schechter function has been used
as follows: $ \phi(L) dL =\phi^{*} (L/L^{*})^{\alpha} \exp(-L/L^{*}) [
1 + (L/L_{t}^{*})^{\beta}] d(L/L^*), $ where L$_{t}$ is the luminosity
at the transition between the two power-laws.  This second power-law
is introduced to fit the overall LF from all galaxies, and in
particular to allow the fit to steepen at the faint luminosities. The
standard Schechter function does not reproduce the up-turn at the
faint-end since most weight in the fits comes from galaxies near
$M^*$.

However if a Schechter function fit is done for each individual galaxy
type, and a final overall fit constructed from the sum of them, this
modified Schechter function is not necessary, since this sum provides
the necessary degrees of freedom to allow a faint-end steepening.  In
fact, various types of galaxies have very different LF; the late types
have a very steep slope with a faint $M^*$, while early types have a
negative slope with a luminous $M^*$.  So the overall LF should always
be calculated as the sum of each individual population LF, i.e.
$\phi(L,{\bf r}) = \sum_{sp} \phi_{sp}(L) \rho_{sp}({\bf
r})/\overline{\rho_{sp}({\bf r})}$ where $sp$ refers to a
sub-population.  These sub-populations can be defined for instance by
colors, surface brightness, morphological parameters (morphology
types, bulge/disk ratios, asymmetry/symmetry, sizes, lumpiness
degrees), spectral parameters (PCA types, line EWs, SED types),
nucleus activity, star-formation rates, their environment, etc.  For
instance, the overall LFs of the preliminary 2dF (\cite{fol}) and of
the LCRS (\cite{bro}) are well fitted over the whole luminosity range
in summing the Schechter function fits of each individual PCA spectral
type, especially at the faint end ($M(b_j)\!-\!5\log h\!>\!-16$) where
the data show a genuine up-turn as in the ESP LF.  On the contrary, a
Schechter function fit for all galaxies does not correctly reproduce
the faint-end slope.

In addition, estimating the LF as the sum of type-dependant LFs avoids
the bad assumption that all galaxies are clustered in the same
way. Measuring the LFs independently is the same as assuming the
separability of $L$ and ${\bf r}$ within each individual population.
However it does not solve the problem if an individual LF depends on
density, as hinted by the results in the LCRS (\cite{bro}).  They find
that the faint-end slope steepens with local density for early-type
galaxies from $\alpha=-0.4\pm0.07$ in high-density regions to
$0.19\pm0.07$ in low-density regions. The strength of such effects is
likely to depend on the classifier chosen to define the
sub-populations.

The last point is that methods which do not make any assumption about
the shape of DF ($\phi$/$\Phi$, STY and SWML estimators), recover only
the shape of the LF (i.e. $\alpha$ and $M^*$), and so must normalize
the LF in an independent manner usually related to an independent
maximum-likelihood estimator. Step-wise estimators calculate the
normalization in each magnitude bins. It is clear that the faint-end
LF reached by a survey is more uncertain due to the small volume
surveyed, and so is more subject to density inhomogeneities.

\section{EVOLUTION OF LUMINOSITY FUNCTIONS}  
A survey with a large baseline in redshift allows the estimation of
the LF at different epochs, and hence allows the detection of
evolution.  This requires observing enough galaxies per bin of
absolute magnitudes in each redshift bin. Although number count
studies suggested evolution in the field galaxy population, it was
controversial (\cite{kron}, \cite{el83}) and only recently has it been
clearly detected observationally (CFRS, LDSS, CNOC2).  Any changes in
the LF with redshift suggest evolution, but care must be taken to
account for incompletenesses and biases, and the significance of any
changes must be compared to the estimated uncertainties.  {\it Any
theoretical interpretation of luminosity functions depends very
critically on an understanding of what is being measured and how it is
measured.}  Parameterizing the LF with a Schechter function adds
complexity, since the three parameters ($\alpha$, L$^{*}$, $\phi^{*}$)
are strongly correlated.  This makes it difficult to disentangle
evolution in density and/or in luminosity.  Moreover if pure density
evolution is detected, it is indistinguishable from density variations
caused by large-scale structure; to infer evolution we must assume
that the universe is homogeneous on very large scales (see
\cite{spri}).

Another critical point is that galaxy populations evolve differently,
and averaging over all galaxies can mask the evolution of each
individual population.  As seen in the previous Section,
sub-populations have very different LF.  The various possible
classifiers are certainly related through star-formation history and
environmental effects, and using several classifiers will allow us to
refine what physically drives the evolution.  Only with large and deep
surveys evolution can be quantified for different galaxy populations.
Moreover selecting galaxies from a single pass-band means that the set
of observed galaxies varies with redshift, and can mimic an
evolutionary trend.  Indeed any particular selection criteria will
favour a particular galaxy population. Deep multi-color redshift
surveys will be better to quantify which set of galaxies is visible at
different redshifts.  Surface brightness functions are also needed to
quantify which galaxies may be missed because of a low (or high)
surface brightness or included due to an enhancement of star
formation.  Studies of low surface-brightness galaxies show that they
are numerous even though they are not a major contributor to the total
luminosity density (see for instance \cite{gau}, \cite{spra},
\cite{lov97}).

Extreme care about the methodologies used should be taken when
comparing faint-end slope from different surveys; a discrepancy can be
mistaken for evolution. The best way to test for evolution is to look
within the same survey and use the same estimators for the comparison,
to avoid the possibles biases discussed earlier in this lecture.  This
point is even more important for LF measured for a particular type of
galaxies, since any classification will be subject to the precise
definition of the classifier which may vary from a survey to another
one, and to systematic variations in the classifier as a function of
redshift.  However to link local surveys to distant ones, we always
have to rely on comparisons between different surveys. Hence the
importance of well-defined surveys.

\section{STATUS}
\begin{figure}
\centerline{\psfig{figure=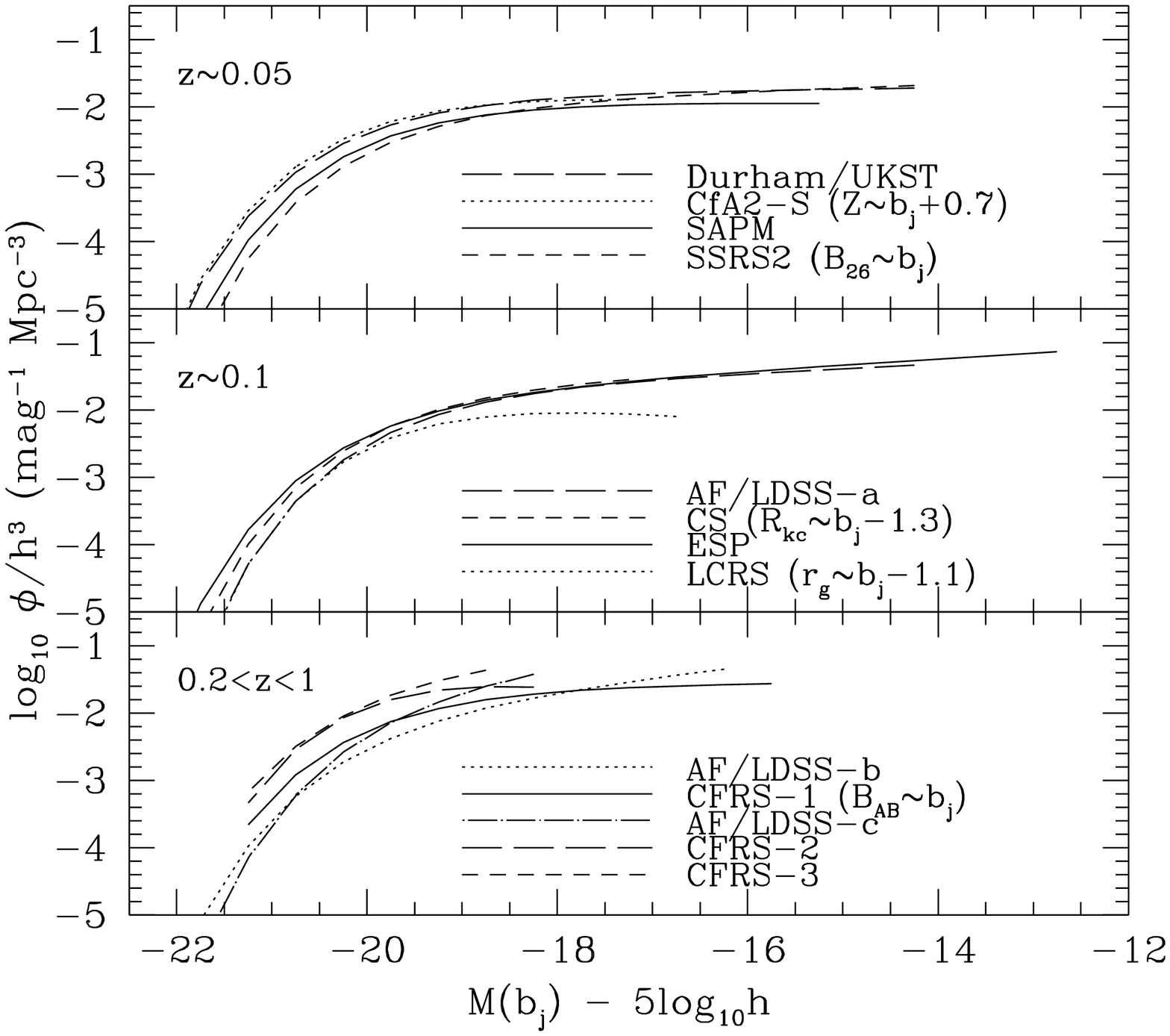,height=10cm}}
\vspace{-1.cm}
\caption[fig2]{LFs with $\bar{z}\!\sim$0.05 (top), with $\bar{z}\!\sim$0.1
(middle), and with $\bar{z}\!>$0.2 (see Table~2). The magnitude conversions 
are taken as quoted in the LF papers.  The LFs are significantly
different between each survey due to the diverse methodologies adopted  
in the selection of galaxies and in the measurements; it emphasizes 
that the quantification of the LF evolution is more securely understood 
within one single well-defined sample. \label{fig2}}
\end{figure}

For 20 years redshift surveys have taken advantage of fast advances in
technology and instrumentation to become larger and deeper.  In the
last few years, the study of galaxy evolution has undergone quite a
revolution, and is still moving fast.  This has been possible thanks
to the development of multi-object spectroscopy, and large sensitive
detectors on telescopes with good seeing. It has led to a much clearer
picture. The qualitative theoretical picture was already in place, but
this has now been refined and quantified by the observational
constraints from recent redshift surveys. Accurate measurements were
crucial to reach the stage where now we can definitely pick the most
likely explanations of galaxy evolution and rule out others that
flourished earlier on.  For the first time, we can trace
observationally the global history of galaxy luminosity density up to
redshifts of $z\!\sim$5 (\cite{mada}). The on-going and future
redshift surveys should represent a major step towards obtaining a
detailed and refined picture of this history.  As we saw, to constrain
better what drives evolution and how galaxies form, we need to
classify galaxies consistently in each redshift bin. We also need to
go deep and far, thus large and deep well-defined redshift surveys are
required.  Table~2 gives the references of papers which estimate LFs,
and Figure~\ref{fig2} displays some of them. Below I give a
non-exhaustive overview of the current status of LFs.
\begin{table}
\caption{Table summarizing published LF analysis from optical-selected
redshift surveys (at the end of year 1998). References give details on
LFs by type of galaxies. Survey annotated by $^\dagger$ are described in
proceedings, so are preliminary results. N$_{gal}$ lists the number of
galaxy redshifts used in the overall LF estimate (usually $\alpha$,
and $M^{*}$), and may differ from the total number of galaxies in the
survey itself).  Selection is the filter in which galaxies have been
selected.  c is for CCD mags, and p for photographic mags (even though some
plate photometry has been re-calibrated later on with some CCD
data). M$_{ref}$ is the pass-band in which LF parameters have been
measured ($h\!=\!H_{0}/100$).}
\vspace{-0.3cm}
\scriptsize
\begin{tabbing}
\hspace{1.6cm}\=\hspace{2.7cm}\=\hspace{0.8cm}\=\hspace{1.cm}\=\hspace{0.8cm}\=\hspace{2.cm}\=\hspace{1.9cm}\=\hspace{1.2cm}\=\hspace{0.6cm}\=\kill
\hrulefill  \\
Survey  \> Selection \> $ \langle z \rangle$ \> N$_{gal}$ \> M$_{ref}$  \> M$_{ref}^*\!-\!5\log_{10}h$ \> $\  \ \  \  \  \  \ \alpha$ \> $\phi^*$x$10^3$/$h^{-3}$ \>   \\
  \> \>  \>  \>   \> \> \>  (Mpc$^{-3}$) \>  Refs.  \\
\hrulefill  \\
CfA2    \> $Z\!\le\!15.5$, p  \> $0.02$  \> 9063 \> Z  \> $-18.75\pm$0.30          \> $-1.00\pm$0.2  \> 40$\pm$10 \>   \cite{mar94}      \\ 
CfA2-N   \>  \>   \> 6312 \>  \> $-18.67$          \> $-1.03$  \> 50$\pm$20 \>   \cite{mar94}      \\ 
CfA2-S           \>   \>     \> 2751 \>    \> $-18.93$          \> $-0.89$     \> 20$\pm$10 \>   \cite{mar94}      \\ 
SSRS2            \> $B(0)\!\le\!15.5 $, p           \>  $0.02$    \> 2919 \> B(0) \> $-19.50\pm$0.8          \> $-1.2\pm$0.7     \> 15$\pm$3  \> \cite{dac} \\ 
            \> $B_{26}\!\le\!15.5 $, p           \>  $0.02$    \> 3288 \> B$_{26}$ \> $-19.45\pm$0.08          \> $-1.16\pm$0.07     \> 10.9$\pm$3  \>  \cite{mar97} \\
           \> $B_{26}\!\le\!15.5 $, p           \> $0.02$    \> 5036 \> B$_{26}$ \> $-19.43\pm$0.06 \> $-1.12\pm$0.05 \> 12.8$\pm$2 \>  \cite{mar98} \\
KOSS \> $F_{KOS}\!\le\!16$, p \> 0.04 \> 229 \> F$_{KOS}$ \> $-21.07\pm$0.3 \> $-1.04\pm$0.30 \> 15.4$\pm$4.9 \>  \cite{efs},\cite{kir1} \\
DARS  \> $11.5\!\le\!b_J\!\le\!17$, p \> 0.04 \> 291 \> b$_J$ \> $-19.56\pm$0.2 \> $-1.04\pm$0.25 \>   $8.3\pm$1.7 \> \cite{efs} \\
SAPM             \>  $15\!\le\!b_J\!\le\!17.15$, p                   \> 0.05 \> 1658 \> b$_J$ \> $-19.50\pm$0.13 \> $-0.97\pm$0.15 \> 14$\pm$1.7  \>  \cite{lov92}  \\
D/UKST \> $b_J\!\le\!17$, p \> 0.05 \> 2055 \> b$_J$ \> $-19.68\pm$0.08 \> $-1.04\pm$0.08 \> 17$\pm$3 \> \cite{rat} \\  
CS               \> $R_{KC}\!\le\!16.13 $, p                  \> $0.08$    \> 1695  \> R$_{KC}$   \> $-20.73\pm$0.18 \> $-1.17\pm$0.19 \> 25$\pm$6.1 \>    \cite{gel}\\
LCRS             \> $15\!\le\!r_g\!\le\!17.7$, p                  \> 0.1   \> 18678 \> r$_g$   \> $-20.29\pm$0.02 \> $-0.70\pm$0.05 \> 19$\pm$1  \>  \cite{lin}  \\
ESP  \>  $b_J\!\le\!19.4$, p \> $0.1$ \> 3342 \>  b$_J$ \> $-19.61\pm$0.07 \> $-1.22\pm$0.07 \>  20$\pm$4 \> \cite{zuc} \\
2dF$^\dagger$   \>  $b_J\! \le \! 20$, p \> 0.1 \> 8182 \> b$_J$ \> $-19.54$ \> $-1.166$ \> 18.3  \> \cite{mad} \\
AF \> $17\!\le\!b_J\!\le\!22$, p  \> 0.15 \> 1026 \>  b$_J$ \> $-19.20\pm$0.3  \> $-1.09\pm$0.1  \>  26$\pm$8 \> \cite{ell},\cite{hey} \\
ESS$^\dagger$  \>  $R_{KC}\!\le\!20.5$, c  \> 0.3 \> 327 \> R$_{KC}$ \> $-21.15\pm$0.19  \>  $-1.23\pm$0.13 \>  $20.3\pm$8 \> \cite{gal}  \\
CNOC2$^\dagger$ \>  $R\!\le\!21.5$, c  \> 0.3   \> 2075 \> B$_{AB}$ \>   $-19.43\pm$0.08  \> $-0.82\pm$0.08  \> $\sim$10  \> \cite{lin98},\cite{car} \\ 
AF/LDSS \>   $17\!\le\!b_J\!\le\!24$, p  \> 0.4 \> 1405 \>  b$_J$ \>  \>   \>  \> \cite{ell},\cite{hey} \\
AF/LDSS-a \>  $\ z$=[$0.02-0.15$]   \>  \> 588 \>   \> $-19.30\pm$0.13 \> $-1.16\pm$0.05   \> $24.5\pm$3 \> \\
AF/LDSS-b \>  $\ z$=[$0.15-0.35$]   \>  \> 665 \>   \> $-19.65\pm$0.11 \> $-1.4\pm$0.11  \> $14.8\pm$3 \> \\
AF/LDSS-c \>  $\ z$=[$0.35-0.75$]   \>  \> 152 \>   \> $-19.38\pm$0.26 \> $-1.45\pm$0.17  \> $35.5\pm$25 \> \\
CFRS \> $17.5\!\le\!I_{AB}\!\le\!22.5$, c  \>  0.56 \>  591 \> B$_{AB}$ \> $-19.68\pm$0.15       \> $-0.89\pm$0.10 \> $32.8\pm$4 \> \cite{lil} \\  
CFRS-1 \>  $\ z$=[$0.2-0.5$] \> \> 208 \> \> $-19.53$ \> $-1.03$ \> 27.2 \> \> \\
CFRS-2 \>  $\ z$=[$0.5-0.75$] \> \> 248 \> \> $-19.32$ \> $-0.50$ \> 62.4 \> \> \\
CFRS-3 \>  $\ z$=[$0.75-1.0$] \> \> 180 \> \> $-19.73$ \> $-1.28$ \> 54.4 \> \> \\
\end{tabbing}
\end{table}
\normalsize

\subsection{Local LFs} 
The local LF still presents questions for debate.  Indeed the
discrepancies between various estimates are not yet fully
explained. However several come from the methodologies and selection
criteria used, so that it is not straightforward to compare the
surveys.  A major issue has been the normalization of the local LFs.
The latest overall LFs at $\bar{z}\!\sim$0.1 of blue-selected surveys
(ESP, CS, 2dFs, AF) are about consistent with a fairly high
normalization and positive faint-end slope.  The LCRS differs for
reasons that are still uncertain, probably due to surface brightness
cuts and/or the selection of galaxies in the red.  At larger depths
$\bar{z}\!\sim$0.3, the ESS, AF/LDSS-b agree also with an high
normalization.  However for less deep surveys $\bar{z}\!\le0.05$,
large discrepancies (up to 50\%) in the normalization estimate are
found between SSRS2, CfA2, SAPM and Durham/UKST surveys.  Multitude of
explanations have been given in the literature, however none has been
fully convincing just because it invokes possible biases in each
surveys that have not been fully quantified yet, while others invoke a
local under-density in the southern hemisphere and/or a rapid
evolution.

The status for local LF estimates can be summarized as follows: \\ 
(a) At $\bar{z}\!<\!0.1$, significant discrepancies are found between
different surveys; the on-going 2dF and SDSS surveys should give more
insights to this problem. \\ 
(b) The LF of blue, strongly star-forming, late-type and/or irregulars
has a steep faint-end slope ($\alpha\!<\!-1$), and faint $M^*$.\\
(c) The LF of red, early-type or E/S0 has a negative slope
($\alpha\!>\!-1$). \\ 
(d) The LF of intermediate-type, spirals has a flat slope
($\alpha\!\simeq\!0$). \\
(e) A faint end cut-off has been yet
not observed at $M(b_j)\!-\!5\log h\!=\!-12.4$. \\ 
(f) LF of galaxies
selected in the optical are steeper than those for a infrared
selection, certainly due to different population sampled. \\ 
(g) Late-type galaxy populations are fainter than early-type ones, and
more numerous at the faint luminosities.   \\ 
(h) An up-turn at $M(b_j)\!-\!5\log h\!<-16$ is genuily detected, 
and is generally related to the low-SB galaxies undergoing significant 
star formation. \\ 
(i) Very late-type galaxies are less clustered than early-type ones.
The strengh of the dependency of the LF with the density for a single
type remains to be defined, however early-type LF seems the most
affected.

These near-UV and blue rest-frame selected LF shapes reflect the
dominant processes in each type of galaxy at low redshifts. Massive
galaxies are not dominated by star bursts, while it is for the faint
galaxy population. Since blue selection is related to the number of
ionizing stars, a steep slope is expected for actively star-forming
galaxies dominated by short-time scale evolution (see
e.g. \cite{hog}).  Knowledge of the shape of these local LFs is
crucial for future distant LFs, to see how each class evolves, and
which ones dominate the evolution at a certain cosmic epoch. Not
discussed in this lecture are the $K$-selected redshift surveys (see
\cite{gar}, \cite{cow}); in this case {\it nearby} galaxies are
selected on their mass even more than with a $R-$ or $I$-selection.
\begin{figure}
\centerline{\psfig{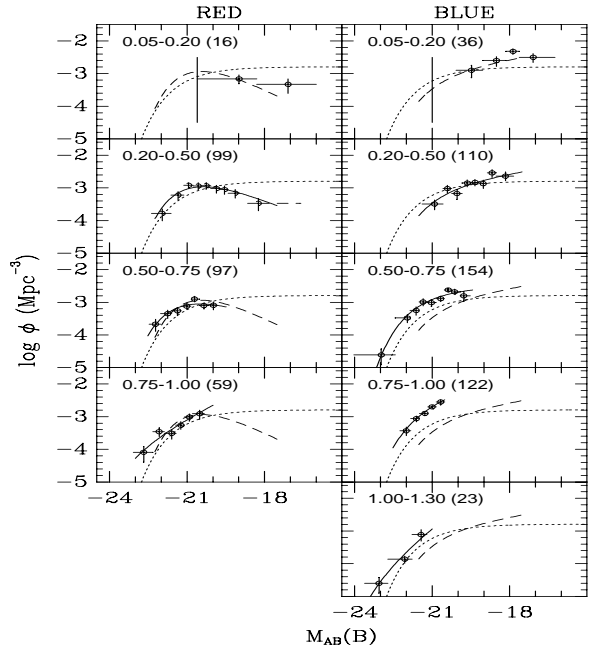}}
\vspace{-1.5cm}
\caption[fig3]{CFRS V$_{max}$ estimates by (V$-$I) color types at
different cosmic epochs, see \cite{lil} for more details
($q_0\!=\!0.5, h\!=\!0.5$). This survey demonstrated definitely the
galaxy population evolution from $z\!\sim$1 to today. \label{fig3}}
\end{figure}
\begin{figure}
\centerline{\psfig{figure=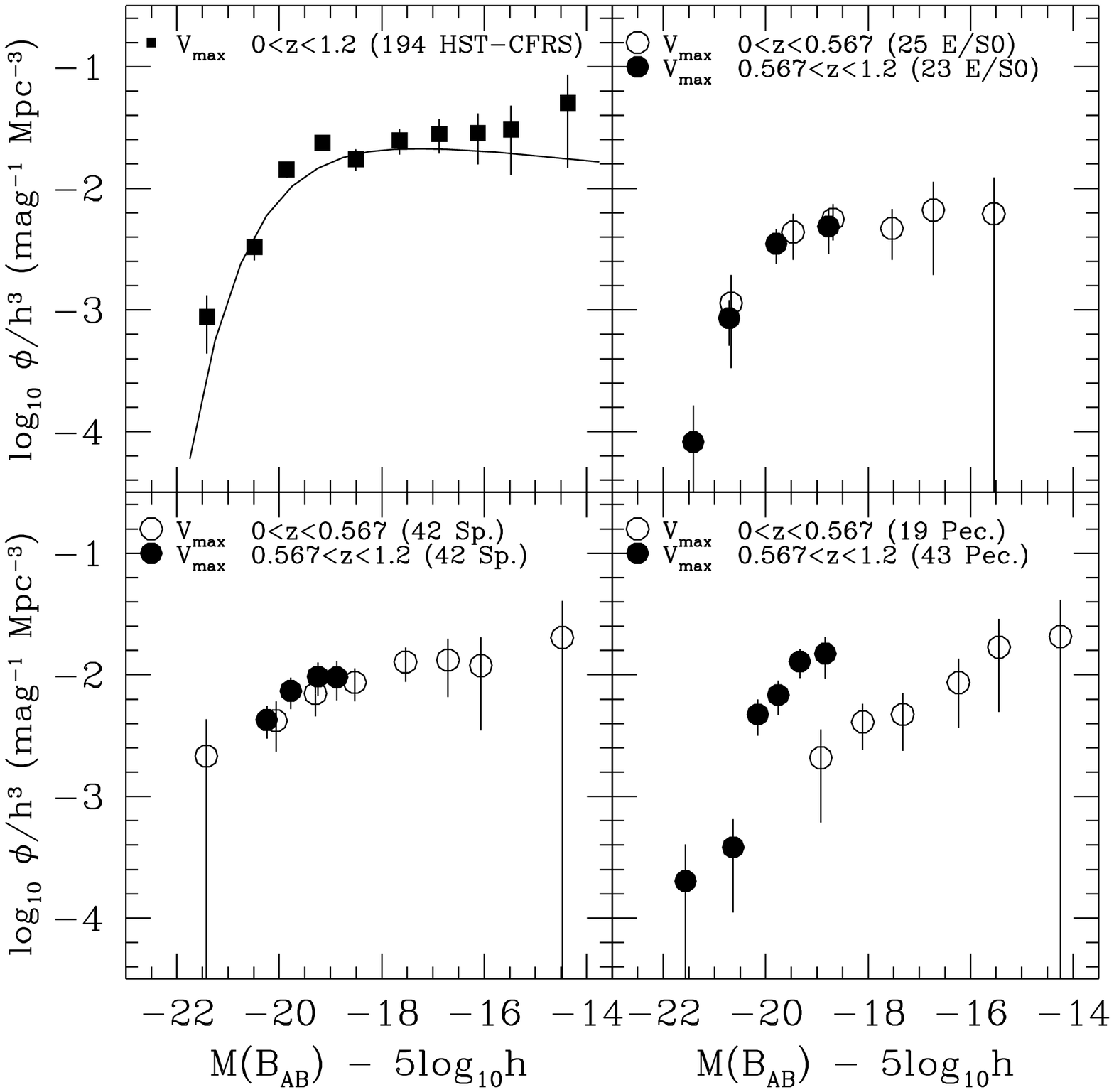,height=10cm}}
\caption[fig4]{V$_{max}$ estimates by HST eyeball morphological types
for a sub-set of the CFRS galaxies. Top-left panel: V$_{max}$ for the
194 HST/CFRS galaxies (dots) and the Schechter function fit of the
whole CFRS (curve).  V$_{max}$ in the low (open dots) and high (filled
dots) redshift range of the CFRS; for E/S0 (top-right), for spiral
(bottom-left) and for peculiar (bottom-right) HST types. Bars are
Poisson errors. ($q_0\!=\!0.5$)
\label{fig4}}

\end{figure}
\subsection{Distant LFs}
The selection of galaxies in the red pass-band allowed for the first
time to probe the evolution of galaxies up to $z\!\sim$1 (CFRS; see
Fig.~\ref{fig3}).  This was followed by the larger CNOC2 survey which
probes more finely the evolution for each individual galaxy type up to
$z\!\sim$0.6. In parallel the Autofib team collected all their
B-selected surveys (DARS, AF-bright, AF-faint, BES, LDSS1, LDSS2), and
also demonstrate evolution in the field population up to $z\!\sim$0.8.

The present status is summarized below, but note that when I write
'compatible', it does not mean that this is the only explanation.
Indeed distinguishing between density and luminosity function requires
further analyses than solely the LF studies. The main conclusions are:
\\ (a) The overall LFs at $\bar{z}\!<\!0.5$ show a normalization
similar to that found at $\bar{z}\!\sim$0.1$-$0.3. \\ (b) Early-type
LF has a negative slope, and evolution is detected for galaxies at
$M(B_{AB})\!-\!5\log h\!<\!-20$ roughly, compatible with luminosity
evolution, and modest or no density evolution. \\ (c)
Intermediate-type LF has a slope almost flat, and evolution is
detected at about $M(B_{AB})\!-\!5\log h\!<\!-20$, compatible with
luminosity evolution. \\ (d) Late-type LF has a steep slope, and
evolution is detected in the steepening of the slope, compatible with
modest luminosity evolution and strong density evolution.  \\ (f) The
faint-end slope at $M(B_{AB})\!-\!5\log h\!>\!-18$ is not yet observed
at $z\!>\!0.5$. \\ The picture of this differential evolution will be
quantified in detail in the near future with larger and deeper
surveys, and several objective classification, and multi-color
selection schemes. The availability of a large well-defined databases
will precisely refine the evolution of each type of galaxy. For
instance we would like to: differentiate between number density
evolution and evolution in clustering properties; to relate
morphological type, spectral properties and environment; to better
constrain the faint-end LF slope at all redshifts for each type of
galaxies. Figure~\ref{fig4} illustrates the V$_{max}$ estimates using HST
eyeball morphological types as tabulated in \cite{bri}; we can see
that the sample is barely large enough to test for {\it overall} LF
evolution split by morphological types. However the estimates around
L$^*$ do agree within the uncertainties with the summary picture
described above.  The absence of bright irregulars at $z\!<\!0.5$ is
particularly noticeable; if they exist they would certainly be
visible. This lack is expected since the L$^*$ of local late-type
galaxies is much fainter than the other types.  The picture at higher
redshifts is significantly different even allowing for possible
misclassification, and is responsible for the strong evolution seen in
deep blue counts (see \cite{bri} for a detailed discussion).
I note that significant discrepancies are likely to be found between
different methods used to define or classify galaxies.  The
acquisition of several colors, and the development of objective
classifications should be a great help to quantify the evolution of
individual population (see e.g. lecture of Abraham for morphological
classification, or \cite{con} for spectral types).  

With this caveat, these deep surveys have shown that the population
which exhibits the strongest LF evolution is composed of galaxies with
strong emission lines, blue colors, peculiar morphologies, and
relatively small sizes.  The LFs for remaining population evolves
mildly or passively mainly at $L\!>\!L^*$. Their faint-end slope is
close to flat, which indicates moderate or little star-formation
activity.  The difference in evolution for these two populations
reflects many different processes in galaxy formation. Some of the
massive systems formed at early epochs ($z\!\ge\!2$), and then have a
declining star formation rate giving the redder galaxies consistent
with passive evolution, while other massive systems still exhibit star
formation at recent epochs.  Smaller systems that form later ($z\sim
1$) are seen during their early phase of intense star formation. Also
smaller gas-rich systems may merge at any $z$; a short starburst phase
during the merger would contribute to the bright late-type galaxy LFs
seen at $z>0.5$, where both irregular morphologies and high
star-formation rate are seen.  A starburst during initial collapse, or
during a merger has a very short timescale, and very quickly a stable
phase is reached where the galaxy fades under passive evolution.
Semi-analytical models and more recently N-body models which
incorporate star-formation prescriptions include all of these
processes, and are giving insights into which are the most dominant
processes in the formation and evolution of galaxies and dark matter
halos (\cite{kau}, \cite{bau}). For a discussion on higher redshifts I
refer to Dickinson's lecture, and to \cite{saw} for the Hubble Deep
Field-North LFs at $1<z<4$. These LFs are preliminary since they are
estimated with photometric redshifts and done on a single field of
view.

\section{CONCLUSION}
Redshift survey analysis has been a crucial step forward in our
understanding of galaxy evolution. We saw that all the steps in making
a survey have their importance and each step needs to be carefully
considered and well-defined.  The luminosity functions are essential
to quantify the evolution for different galaxy types.  Larger and
deeper surveys are necessary to precise the differential evolution of
galaxy populations at all redshifts. Local surveys will give a refined
and clearer picture at $z<0.2$.  Future surveys on 8m class telescopes
equipped with new infrared capabilities will allow us to refine the
evolution at $z\!<\!1$, and systematic redshift surveys at $z\!\gg\!1$
will be enabled. At the same time deep redshift surveys in other
wavelengths (UV, far-IR, mm, radio) are also crucial to link the
different emissivities of the galaxy populations, and to observe
epochs of formation of the first stars. These will be crucial to
differentiate between models of galaxy formation and evolution.\\

ACKNOWLEGMENTS I thanks J. Loveday and S. Maddox for their careful
reading of the lecture. I thanks the organizers O. Le F\`evre and
S. Charlot for this enjoyable school and for providing financial
support.

\end{document}